# A Neural Networks Model of the Venezuelan Economy

Sabatino Costanzo, Loren Trigo, Luis Jiménez, Juan González

**Abstract**

Besides an indicator of the Gross Domestic Product, the Central Bank of Venezuela generates on a monthly basis the Monthly Economic Activity General Indicator (IGAEM). The a priori knowledge of this indicator --which represents and sometimes even anticipates the economy's fluctuations--, could be helpful in developing public policies and in investment decision making. The purpose of this study is forecasting the IGAEM through the use of non parametric methods, an approach that has been proven effective in a wide variety of problems related to economic and financial analysis. When defining the problem of forecasting the Venezuelan economy, the IGAEM was selected as the variable to be predicted. Historical data for this indicator was available from 1991 to 2003. Such data was divided into training and testing ranges. The testing data was used to ascertain the predictive value of the neuronal networks, or *experts*, as we will call them from now on. The training period went form January 1992 until December 1999 and the testing period from January 2000 until December 2003, according to the pre-processing rules of the model's variables that will be described further along. Useful pragmatic results were obtained from this study. The forecasting system consisting of a master network with eight sub networks, yielded a hit rate of over 80%, suggesting the presence of discernible patters in the data. There is evidence suggesting that neural networks could be more efficient working in groups than individually. In fact, our master networks present efficiency indicators –Sharpe Ratio and hit rate— that are way above those of the sub networks. In a subsequent study it would be interesting to try to outperform the actual results taking into account alternative lags in the input variables.

# 1. INTRODUCTION

In the 20's decade, as the industrial exploitation of oil in Venezuela began --and oil income became more and more influential in the Nation´s future--, the dependency of the economy on oil-mining became patent in the composition of the Venezuelan GDP.

The GDP is structured in two segments: "Oil Industry related GDP" and "Other activities related GDP".

The Oil Industry could be seen as an "economic growth engine" because its demand of goods and services from the rest of the economy has an indirect impact on the non-oil related aspect of the GDP that is even larger than its direct contribution.

The growth of the Venezuelan economy has been always conditioned by boosts in the production capacity and by fluctuations in oil prices directly affecting public expenditure planning. Such fluctuations generate a surplus in the fiscal accounts during price upswings and a deficit during downfalls, affecting the GDP's stability by transferring onto it the volatility inherent to the nature of oil prices. Other sources of uncertainty such as political crises, variations in interest rates, inflation, devaluation, capital controls and economic measures, contribute -- by directly affecting the GDP--, to an unstable environment which is not appropriate for sustainable investments.

It is the Central Bank's duty to capture the essential elements that are present in the reports of the different sectors of the economy in order to generate, on a monthly basis, the Monthly Economic Activity General Index (IGAEM), a good monthly estimator of the GDP. It is obvious that acknowledging a priori the GDP's fluctuations by doing it with the LEI's movements would be decisive when developing public policies and in investment decision making. The forecasting of the IGAEM has been unsuccessfully attempted many times before through multiple linear regressions. The purpose of our study is to address the same problem from a non parametric approach which has proven to be effective in predicting non-linear processes going from credit authorizations, portfolio selection and mortgage risk analysis, up to market behavior simulation, index construction and economic indicator movements prediction.

## 2. DEFINITIONS

### 2.1 THE IGAEM

The Monthly Economic Activity General Index (IGAEM) is a periodic index geared towards evaluating the economy's evolution in the short term, based on a set of monthly indicators elaborated by the Central Bank of Venezuela (BCV).

In order to construct the index, the BCV selects a set of relevant indicators reflecting most of the economic activity. Initially, all of the sectors in which the GDP is disaggregated are included, and later on, additional indicators linked to the behavior of each one of these, are added.

Since some sectors have no intra-annual indicators --or these indicators do not represent the sector being studied--, the BCV excludes them from the IGAEM's calculation, the reason why the IGAEM is limited to a subset of sectors not including all of the activities that conform the GDP. (Even though for example, the added value reached for the base year (1997) is 80% of the total national economic activity). The IGAEM is an index of the Laspeyres kind, that is, a measure with fixed weightings, in one base year, of the relative quantities that constitute it. The general index's weights correspond to the participation of each one of the GDP's activities for the base year. A table containing these weights is shown next:

**Table 1: Weights of the Productive Sectors Within the GDP, Base 1997**

| Activity | Weights |
| --- | --- |
| Oil | 20.9 |
| Mining | 0.8 |
| Private Manufacturing | 10.5 |
| Water and Power | 1.5 |
| Construction | 5.2 |
| Commerce | 11.7 |
| Financial Institutions and Insurance Companies | 2.3 |
| Real Estate | 6.8 |
| Professional Services | 3.3 |
| Communal, Social and Personal Services | 9.6 |
| Imports Rights | 7.4 |
| **Total** | **80.0** |

In order to evaluate this indicator as a good estimator of the GDP, the following table shows the proportions in which the IGAEM and the GDP have the same direction of change:

**Table 2: IGAEM's Hit Rate versus GDP**

|  | 1985.1-1999.2 | 1990.4-1999.2 | 1996.1-1999.2 |
|---|---|---|---|
| **Hits** | 54 | 35 | 13 |
| **Number of Observations** | 57 | 35 | 13 |
| **Percentage** | 95 | 100 | 100 |

In order to evaluate the magnitude divergences between the GDP and the indicator, two error measures were calculated between the inter-quarterly variations of the index and the GDP's variations.

a. Mean absolute error: $\sum_{i=1}^{n} \frac{|PIB - IGAEM|}{n}$

b. Mean quadratic error: $\sqrt{\frac{1}{n} \sum_{i=1}^{n} (PIB - IGAEM)^2}$

**Table 3: Estimation Errors**

| Quarter | Mean Absolute Error | Mean Quadratic Error |
|---|---|---|
| **1985.1-1999.2** | 3.0 | 3.9 |
| **1990.4-1999.2** | 2.3 | 2.7 |
| **1996.1-1999.2** | 1.9 | 2.3 |
| **1998.1-1999.2** | 1.5 | 2.0 |

Just as shown in the previous table, the errors keep lessening as the sample becomes contemporary, which could be explained by the incorporation of methodological improvements in the construction of the indicators used by the BCV to calculate the IGAEM.

Thus, the IGAEM seems to be an excellent indicator of the economic reality and a reasonable predictor of the GDP's variations both in direction (expansion or contraction) and in magnitude, such as the error measures suggest.

**2.2 NEURAL NETWORKS**

Before addressing the problem of the methodology and data used in this study, a general idea about the definition of a neural network should be given, and in particular, of the characteristics used in this study as well as their training.

We will focus our analysis in multi-layer neural networks with supervised learning because, as we will see further along, these are capable of adjusting to any function and perform particularly well with time series.

As in other projection models such as regressions and extrapolations (e.g. moving averages, smoothening), models based on neural networks use inputs in order to generate a result which is generally an estimate or a projection.

What distinguishes neural models from others is their ability to learn and adapt themselves to the environment.

Neural networks are composed of three basic elements:

1. Processing units or nodes working in parallel.
2. Transfer functions (or activation functions) that transform the nodes' information.
3. Connection weights that determine the relative importance of nodes.

In the construction of a network there can be one or more layers. In the case of the feedforward multi-layer networks (MLP or Multi Layer Perceptron), one layer's outputs constitute an input for the next one, as shown in the following graphic:

*A three layer Multi-Layer Network has the following form:*

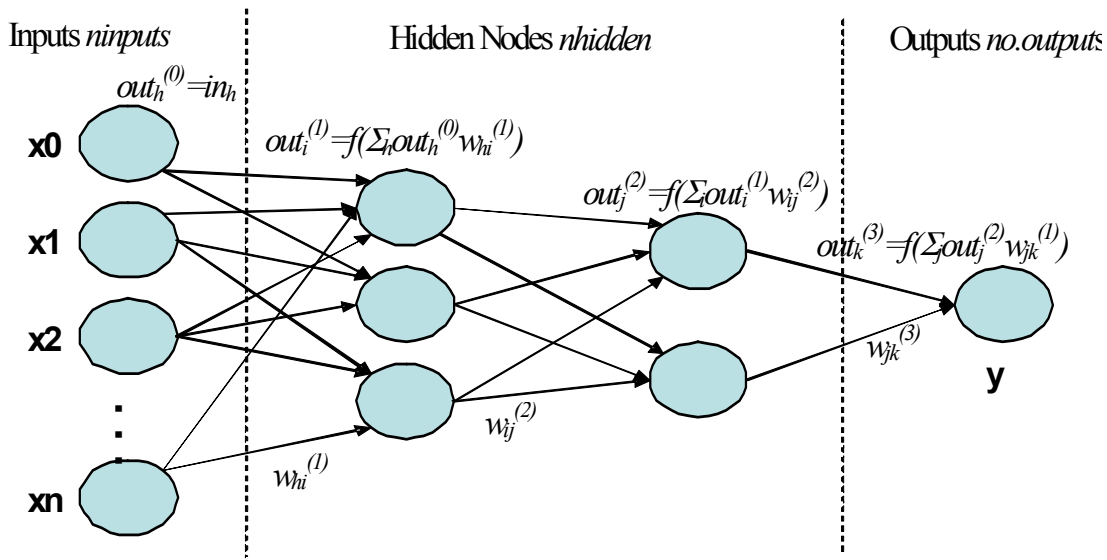

For convenience, when referring to a network of *n* layers, the input layer is layer 0 and the output layer is layer n. The mid layer(s) is called, for our convenience, hidden layer(s).

The neural networks learn and adapt themselves by modifying the weights associated to the connections between nodes.

Some important factors involved in the learning of multi-layer networks are the following:

1. The purpose of the network's training or learning process consists in minimizing the output errors over a particular training data set by updating the connecting weights *wij*. The rules that govern the updating process of the weights constitute the network's learning algorithm.
2. We defined an Error Function *E(wij)* that "measures" how far the actual network is from the desired network (the properly trained network).
3. The partial derivatives of the error function ( d*E(wij)* / d*wij* ) indicate in which direction we should move within the weights range in order to reduce the error. This component of the learning algorithm is called Descending Gradient.
4. The learning rate *h* specifies the size of the steps to be taken in the weight space for each iteration of the Descending Gradient equation updating such weights.
5. Only the outputs of the final layer appear on the Error Function. However, this error will depend on all the layers of previous weights, and a component of the learning algorithm called Back Propagation will adjust them all. The Back Propagation automatically adjusts the output of the previous hidden layers in such a way that the layers form (hidden) appropriate intermediate representations.
6. We continue step by step through the weights space until the errors are "small enough".
7. If we choose neuronal activation functions (transference) with derivatives that take particularly simple forms, we can make the counts used in weight updating, very efficient.

**Training Multi-Layer Networks (MLP)**

1. We choose the set of training patterns that we want the network to learn $\{in_i^p, out_j^p : i = 1\ldots ninputs, j = 1\ldots noutputs, p = 1\ldots npatterns\}$.
2. We configure the network with input units (*ninputs*) completely connected to hidden units (*nhidden*) through connecting weights *wij*, which at the same time are completely connected to output units (*noutputs*) through connections with *wjk* weights.
3. We generate random initial weights, e.g. of range [–*smwt*, +*smwt*].
4. We select an appropriate Error Function $E(wjk)$ and learning rate.
5. We execute the weight updating process $Dwij = -h \P E(wij)/\P wij$ for each *wij* weight, for each *p* training pattern. A set of updates of all the weights for each of the training patterns are called a training period.
6. We repeat step 5 until the network's error function is "small enough".

This is how we obtain a trained neural network.

As it can be deduced from above, training a network involves defining a significant amount of parameters which opens the possibility of an over parameterization or over fitting of the model to the data. There is a technique that optimizes neural network training avoiding the over adjustment that we will present at the end of this section. To appreciate it we have to understand how this over fitting occurs, which we will do next.

**Computational Power of Multi-Layer Networks**

According to the theorem of universal approximation for MLP's proven independently by Cybenko and Hornik (1989), any continuous function capable of *mapping* real number intervals to some real number *output*-interval, can be approximated in an arbitrarily close way by an MLP with just one hidden layer. This result holds true only for restricted kinds of activation or transference functions such as sigmoid or logistical distribution functions. In other words, j(x) is a growing-monotonous, continuous, non-constant and delineated function. Then, for any continuous function f(x) with x={$x_i \in [0,1]$ : i = 1, …,m} and e > 0, there is an integer M and real constants { $a_j$, $b_j$, $w_{jk}$ : j = 1, …,M , k = 1, …,m } such that $F(x_1 \ldots x_m) = \sum_{j=m}^{M} \alpha_j \varphi \left( \sum_{k=1}^{m} w_{jk} x_k - b_j \right)$ is an approximation of f(x), that is, $\left| F(x_{1,\ldots,} x_m) - f(x_{1,\ldots,} x_m) \right| < \varepsilon$ for every x that lies in the input space. Clearly, this is applicable to a MLP with M hidden units because j(x) can be a sigmoid, $w_{jk}$, $b_j$, can be hidden layers and biases, and the $a_j$ can be output weights. From here it follows that given enough hidden units, a two-layer MLP can approximate any continuous function. For a statistical interpretation of neural networks we recommend that the reader should refers to the book of Bishop (1995).

**2.3 EFFICIENCY INDICATORS EMPLOYED**

Because there are so many aspects in which two time series can be compared, such as the IGAEM and the indicators that will be used in order to predict it, and to the fact that it is not possible to sum all of them up in only one indicator, it was necessary to resort to a series of them in order to proceed with the comparison. Some of these indicators are:

- **Hits:** They indicate the percentage of hits in which the predicting series coincides (in terms of direction) with the real series, both in ascents and descents. This way, a predicting series that obtains a 50% hit rate means it coincided with the historical series the same number of times in which it didn't:

- **Efficiency:** The efficiency tries to quantify the profit margin that would be gained it the time series were marketable. This means treating the time series as if it were prices of a certain stock that can be bought or sold in a stock market and one in which a monetary payoff can be obtained, buying or selling the series according to the upswing or downswing tendency. The maximum profit is defined as if the stock was accurately negotiated a 100% of the times. The profit obtained with the predicting series will always be less or equal to the maximum profit. This percentage indicator is given by the quotient of the realized gains over the maximum gains obtainable multiplied by 100.

- **Mean Error:** This indicator measures the magnitude of the average difference between the predicting series and the real series and is expressed in the same units as the series being studied. A low mean error indicates that the predicting series is, in average, very close to the real series: EAM is nothing else but the quotient of the summation of the n differences occurring among the estimate and real values, all divided by n, and EAM´ is the same average but with the differences squared.

- **Mean Quadratic Error:** As with the previous indicator, this indicator evaluates the magnitude of the difference between the predicting and the real series, with the exception that both positive and negative difference are given the same treatment, and so avoiding that positive and negative errors counteract themselves. This measure gives a better view of how close two series are in terms of magnitude: ECM is the square root of EAM´.

- **Modified Sharpe Ratio:** This indicator has the particularity of achieving a balance between the efficiency measure of the predicting series and its consistency in order to achieve that efficiency. Two networks that have an 85% efficiency aren't equally good if one of them achieves it with less volatility in its losses. The less volatile series has a lower degree of uncertainty and therefore will have higher Modified Sharpe Ratio. If $k$ is the number of losses during the total number of $n$ events: SRM is the quotient of Efficiency over the average draw-down

3. **PREDICTION MODEL DESCRIPTION**

3.1 **ARCHITECTURE**

The proposed system consists of eight neural networks, each one of which seeks to predict the Venezuelan economy indicator IGAEM by using different inputs. The result that each one of them yields is of considerable predictive power, as is evidenced in the results analysis section.

Another network which we will denominate Master Network uses the outputs of the eight networks initially obtained as its inputs, and the IGAEM as its output.

Next, a diagram of the complete system is shown:

**Figure 5: General Diagram of the System**

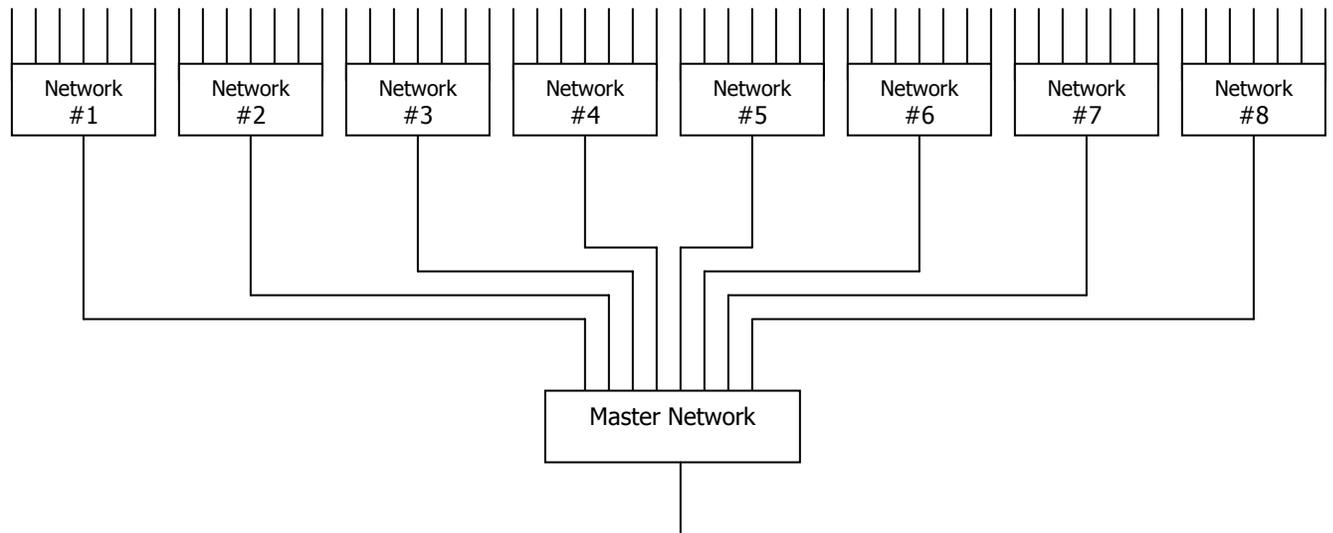

## 3.2 MODEL'S INPUTS

The gathered data has monthly frequency and comprises a 13 year period, from January 1991 to December 2003. A longer time frame was not considered due to non availability of IGAEM data. We can state that oil prices are considered as inputs for the models given their direct and indirect influence over the Venezuelan economy's fluctuations; the existence of a series of historically proven correlations between commodities, currencies and financial ratios revealed by the Intermarket Analysis suggest that these could add valuable information to the predictive model, and for reasons similar to the previous ones, it is also important to include variables such as power consumption, the stock index and inflation, which presumably also contain relevant information to our purpose.

*Brief Description of the considered Inputs:*

- GWH: Power Consumption measured in Giga watt hour. Information source: National Interconnected System Planning Office (OPSIS).
- IBC: Caracas Stock Index, expressed in December 1993 basic points. Information Source: Caracas Stock Exchange.
- Loan rate: monthly bulletins issued by the BCV, website: **www.bcv.org.ve,** expressed in nominal terms.
- Light Crude: Source: *Pifin, ticker* CL.
- IPC: consumer price index (percent variations). Source: BCV.
- S&P500: Standard and Poor 500 Composite Index. Yahoo Finance.

- T-Bills: interest rate paid by 90 day T-bills. Source: *Pifin*.
- Gold: 100 oz. price series. Source: *Pifin*.
- Copper: copper price series. Source: *Pifin*.
- Eurodollar: rate of return of the American dollar in Europe. Source: *Pifin*.
- CRB: commodities index generated by the *Commodities Research Bureau*. Source: *Pifin*.
- Dow Jones Utility: index of the United States' main utilities companies. Source: *Pifin*.

## 3.3 DATA PREPROCESSING

There is a series of preprocessing techniques that ease the net's learning process and pattern attainment. These techniques range from "smoothing" the data's fluctuations and peaks through *Moving and Block Averages* to converting non-stationary series into stationary ones. Albeit the neural network gifted with back percolation is less sensitive to the data's intrinsic imperfections than econometric models in which the inputs' character and behavior are key, all of the data with which the networks have been fed have been previously preprocessed with the objective of avoiding receiving duplicated or contradictory information. For example, the usage of *moving averages* in order to filter the chaotic noise in everyday prices leaving only one soft tendency line, is a procedure motivated by the fact that a neural network performs better predictions when the output is not a sequence of tiered data. If conventional moving averages (MA) allow smoothening a temporal series by eliminating the chaotic noise of everyday prices by substituting them for a soft line, the *block averages* (BA), similar to the conventional MA's except these are taken $n$ distance units away from a $t$ instant used as reference, are used to identify cyclic patterns like the "lags", while the variations time series obtained from the [ Observation $_i$ – Observation $_{i-1}$ ] differences allows, in our case, counteracting in a simple way the non-seasonality of the most problematic portions of the time series.

### 3.3.1 OPTIMAL LAGS DETECTION

If each input were evaluated on its ability to predict the IGAEM, it would become evident that, as the input lags itself from one month to another, the results –in terms of predictive ability— vary sensibly. In order to determine which of the lags of each entry is the optimal in terms of the prediction, a program was designed that estimates the fluctuations of the predictive ability of each lag, evaluating it as if it were being used as an forward indicator of the economy's change of direction

(upswing / downswing) –this evaluation was effectuated through a battery of indicators (see section 2.3) to indicate which lag yields the best result--.

In Figure 6, the curve with the highest slope (dark blue) represents the "*perfect equity*", which is the accrued profits curve (graphic of a virtual "bank account") that could be obtained if the IGAEM were a negotiable "Commodity" and we would have been right on the predictions of <u>all</u> the upswing or downswing movements of the curve, buying on every upswing and selling on every downswing, with no exception. This *benchmark* works as an *upper bound*.

The curve in fuchsia represents the outcome of buying a stock and keeping it for 4 years, which corresponds to the normal movement of the economy: when this one rises, profits rise, and vice versa. This *benchmark* acts as a *lower bound*. The 12 remaining curves represent the 12 lags. As evidenced, not every lag offers the same final profit and not all of them have similar slopes on every evaluated point throughout time. The optimal lag will not only be the one that produces the most profits, but the one that achieves it in the most consistent way, that is, without abrupt or deep upswings or downswings, which is equivalent to having the maximum Modified Sharpe Ratio. As it is shown on the same figure, the dark green lag (representing the twelve months lag) clearly offers the highest profit and has a uniform slope throughout the test period. This will be the lag used as an input for the selected networks.

Table 5 shows a summary of the optimal lags obtained for each input as well as the input structure assigned to each of the 8 networks comprising the Master Network.

**Table 5. Inputs for each Network**

| Input | Network 1 Beta 0,2 | Network 2 | Network 3 Beta 0,25 | Network 4 | Network 5 | Network 6 Beta 0,1 | Network 7 Beta 0,2 | Network 8 Beta 0,3 |
|---|---|---|---|---|---|---|---|---|
| GWh (lag –10) | | | X | | | | | |
| IBC (lag –5) | | | X | | | | | |
| S&P 500 (lag –5) | | | X | | | | | |
| Light Crude (-7) | | | X | | | | | |
| T-Bills (lag –10) | | | X | | | | | |
| Gold (lag –2) | | | X | | | | | |
| Copper (lag –8) | | | X | | | | | |
| Eurodollar (lag –10) | | | X | | | | | |
| CRB (lag –12) | | | X | | | | | |
| Dow Jones (lag –3) | | | X | | | | | |
| Loans Rate (lag -1) | | | X | | | | | |
| IPC (lag –9) | | X | X | X | X | | X | |
| Var GWh | | X | | X | X | | X | |
| Var IBC | | X | | X | X | | X | |
| Var S&P 500 | | X | | | X | | X | |
| Light Crude VaR | | X | | X | X | | X | |
| Var T-Bills | | X | | | X | | X | |
| Gold VaR | | X | | | | | X | |
| Copper VaR | | X | | | X | | X | |
| Eurodollar VaR | | X | | X | X | | X | |

| | | | | | | | |
|---|---|---|---|---|---|---|---|
| CRB VaR | | X | | | X | | X | |
| Dow Jones VaR | | X | | | | | X | |
| Loans Rate VaR | | X | | X | X | | X | |
| MA (Igaem) (lag–12) | | | X | | | | | |
| Log (Var(MA(t))) Lag -12 | | | | X | | | | |
| Standard Deviation | | | | X | | | | |
| 5 MA's Lags Fourier series | X | | | | | X | X | X |
| 4 BA's Lags with Fourier series | X | | | | | X | X | X |

## 4. METHODOLOGY: OPTIMIZING THE GENERATED NEURAL NETWORKS TO PREDICT THE IGAEM

Creating a neural network that yields good results requires an important trial and error process during which it is important to follow certain empiric rules that increase the network's ability to identify patterns from the data set that feeds it.

In order to model the problem of predicting the IGAEM it was necessary, after defining the input and output variables, to collect 13 years of monthly historical data for the output (IGAEM), as well as for each one of the inputs previously described. The data was distributed in two categories: one for training and the other for testing which would serve to ascertain the *expert's* predictive power. The proportion of data used for training was 66% versus 33% used for testing. Some input variables proved having up to a 12 month optimal lag with respect to the IGAEM, being of the outmost importance when studying the cycles detected in a one year period through the *moving* and *block averages*. The training period went from January 1992 until December 1999, and the testing period, from January 2000 to December 2003, saving the year 1991 for the preprocessing adjustment of the model's variables.

After defining the training and testing periods, an expert was created from each one of the networks that were originally generated after focusing on their architecture's optimization and the minimization of their estimation error, looking for maximizing the Sharpe Ratio.

**Architecture Optimization**

This process consists in finding the expert's configuration that best adjusts to the number of variables and available data, which means finding, through the usage of a proprietary algorithm, the number of hidden layers and the number of nodes on each hidden layer that minimizes the error percentage when processing both training and testing data. Besides its configuration and the number of inputs and outputs of the neural network, it is important to determine the kind of data to be predicted. Since IGAEM is the variable to be predicted, the algorithm's "*linear*" mode was selected as it is the one that best models the outputs –which can be represented as continuous-- among the other possible generated values.

The *Initial Weight Ranges* is the algorithm's feature that determines the values range in which the expert will assign random numbers to the connection weights between the nodes each time that the network is trained, while the Learning Rate modifies the network's tendency to change the connection weights, forcing it to make progressively smaller changes for each new repetition of the recursive process. The described recursive process ("S*earch for Best Net*") was used for optimizing both the sub networks and the Master Network.

**Maximizing the Sharpe Ratio**

This process, executed through a Visual Basic program whose function consists in recursively retraining the expert until achieving the desired Extended Sharpe Ratio, consists in randomly regenerating the nodes' connection weights on each iteration. On this Darwinian elimination process, only the most apt neural network survives. The Sharpe Ratio Maximization was used in the sub networks --as well as in the Master Network--, after having optimized the architecture and making a parallel observation of other indicators such as the hit rate, error percentage in the training data range and the error percentage in the testing data range.

## 5. ANALYSIS OF THE RESULTS

Table 5 shows that networks 1 and 5 have the same hit rate, being third in the networks LIST. However, the efficiency percentage of network 5 is less than it is for network 1, which is consistent with the Sharpe Ratios obtained, given that even though they have the same number of hits, network 5 misses in positions that generate a higher loss and therefore increase negative volatility. It is interesting to point out that networks 1 and 5 have similar *block averages* and *moving averages´* inputs with different Betas --the first one with a Beta=0,2 and the second with a Beta=0,3--, which suggests that the network learns to make better decisions as the *moving average* curve becomes softer. Meanwhile, network 5 shows higher magnitude errors than network 1, as it is evident in the training and testing error percentages.

**Table 6. Indicator's Results of all the networks**

| Networks | Efficiency % | Hit % | Sharpe Ratio | Mean Quadratic Error | Mean Error | Training Error % | Testing Error % |
|---|---|---|---|---|---|---|---|
| Network 1 | 60,21% | 70,31% | 0,6496 | 8,11 | 2,67 | 14,72% | 19,41% |
| Network 2 | 48,31% | 64,06% | 0,4011 | 11,72 | 2,47 | 19,54% | 34,58% |
| Network 3 | 52,66% | 64,06% | 0,4775 | 9,14 | 1,89 | 14,30% | 25,95% |
| Network 4 | 42,54% | 76,56% | 0,3178 | 10,61 | 1,64 | 17,34% | 31,30% |
| Network 5 | 45,19% | 70,31% | 0,3539 | 11,41 | 3,90 | 16,55% | 32,99% |
| Network 6 | 49,55% | 68,75% | 0,4216 | 8,00 | 2,76 | 13,81% | 19,96% |
| Network 7 | 63,61% | 71,88% | 0,7501 | 8,95 | 3,50 | 12,72% | 23,23% |
| Network 8 | 51,80% | 68,75% | 0,4613 | 9,15 | 3,67 | 15,57% | 22,13% |
| Master Network | 76,18% | 75,00% | 1,3724 | 7,34 | 2,41 | 11,68% | 19,78% |

The sub networks with the highest Sharpe Ratios are 1 and 7, with network 7 outperforming network 1, but both with hits slightly higher than 70%. The efficiency percentage shows that network 7 hits increased performance, even though the magnitude errors in *testing* were higher than those of network 1. In fact, the error percentage of network 7 is 12,72%, much lower than its testing error percentage (23,23%), which could indicate that the network is over trained.

Networks 4 and 7 have the best hit rate; however, 7 has a higher Sharpe Ratio and lower error percentages than 4, with market variables as inputs and *block averages* and *moving averages* with Beta=0,2 while network 4 only uses market variables, a single *moving average* and the accrued standard deviation.

Another interesting fact is that networks 2 and 3 have the same hit rate even though 3 has a better Sharpe ratio than 2. Furthermore, the inputs for both networks are similar: network 2 uses transformed market variables, while network 3 uses the same market variables without transformations and a weighted moving average with a Beta=0,25 and a 12 month lag, which suggests a strong annual cyclical component in the Venezuelan economy.

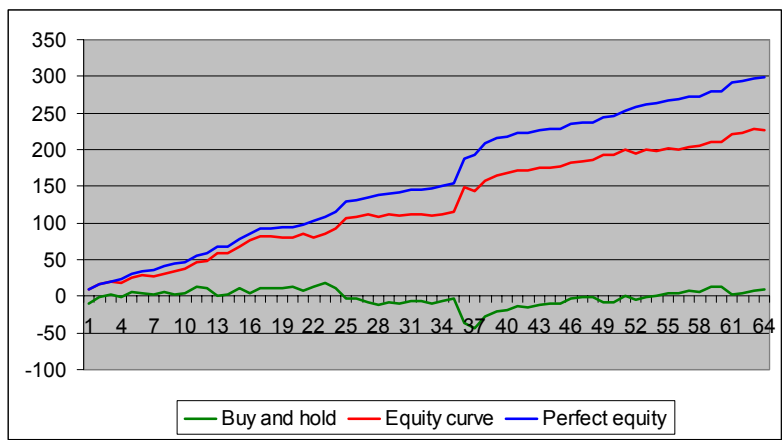

In figure 7, the Master Network's *Equity Curve* significantly adjusts itself to its *Perfect Equity*, placing itself way above the *Buy and Hold*.

**Figure 7: Prediction's Equity Curve**

On the other hand, notice that the networks showing the highest mean quadratic error are 2, 4 and 5. None of these networks got right estimating the economy's downswing magnitude as a consequence of the strike that took place on December 2002, while networks 1 and 8 did. The former have *moving averages* and *block averages* with 4, 5, 7 and 9 month lags as inputs. Networks 2 and

4 only use inter-market analysis variables, and networks 3 and 5 incorporate a *weighted moving average* with a 12 month lag.

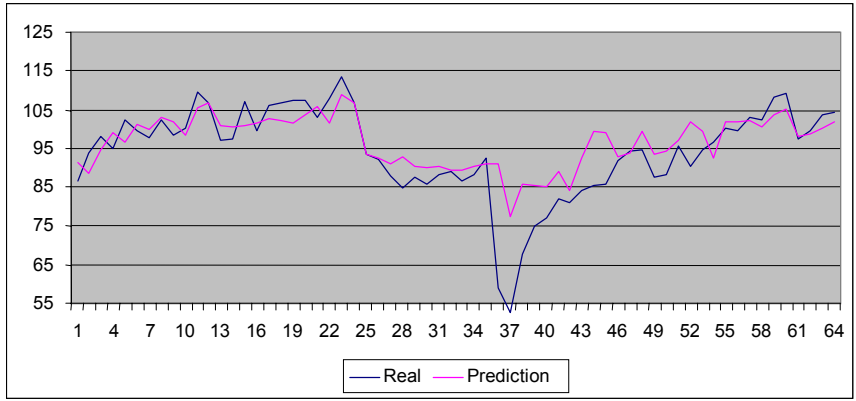

Figure 8 shows the behaviors of the IGAEM's Prediction Curve vs. the IGAEM's Real Curve.

**Figure 8. IGAEM's Curves vs. Prediction**

On this example, it is evident that a significant portion of the mean quadratic error seems to obey to the economy's downfall of 2002 as an atypical phenomenon which the network is still capable of detecting. It can also be deduced from the previous graphic that the mean error is high in comparison to some of the sub networks, since most of the time it remains above the real curve while the economy recovers.

**CONCLUSIONS**

This research has shown practical results. The base of the prediction system conformed by a master network with eight sub networks yield a hit rate higher than 70%, which suggests the existence of patterns with cyclical character.

On the other hand, the *moving averages* (MA's) seem to represent a determining factor in the mean quadratic error's reduction (ECM) because these allow the prediction series to better adjusts itself to the real series. Only the sub networks with Fourier decomposition, MA's and BA's, seem to be able to adapt themselves to the IGAEM's downfall, as a consequence of the late 2002 strike, and they are precisely the ones that present the lowest ECM's.

Similarly, neural networks working together have a much higher predictive power than when they work individually. The Master Network shows efficiency indicators, Sharpe Ratio and a hit rate way above those of the sub networks. Even more, both the *testing* error and the ECM are the lowest observed.

In general, it has been shown that these networks could be used in investment decision making to know in advance whether the economy will be expanding or contracting, allowing to prepare for possible rises in a product's demand, anticipating a recession or taking actions over aspects such as inventory variations, working capital variations and execution of marketing plans.

Finally, it should be pointed out that it is possible to take advantage of these neural network's features in order to acknowledge the relative importance of inputs in the economy's prediction and that its predictive capacity can be preserved throughout time by using a simple algorithm of periodic retraining.

**6. FUTURE RESEARCH**

The largest anticipation with which a neural network can predict in a way depends on the input variable's lags. In this study, a model capable of predicting four months in advance was obtained. However, it would be interesting to improve on these results by considering a larger lag, as well as finding an input variable that somehow reflects the political and economical environment, able to forecast the magnitude of the economy's abrupt downfall, consequence of the strike that took place in late 2002.

Similarly, it would be of interest to use the Master Network for calculating a daily synthetic index and applying statistical tests to it in order to determine a beta that reflects the Venezuelan market's volatility in comparison with U.S.'s, framed in the general theory of the *Capital Asset Pricing Model* (CAPM) valuation model.

Also, other data analysis tools (*Data Mining*) such as *Rough Set Engines* based on *Rough Set Theory*, could be used with the neural networks in order to refine the results that obtained in this study.

## 7. BIBLIOGRAPHIC SOURCES

**APPENDIX**

**Network 1 Equity Curve**

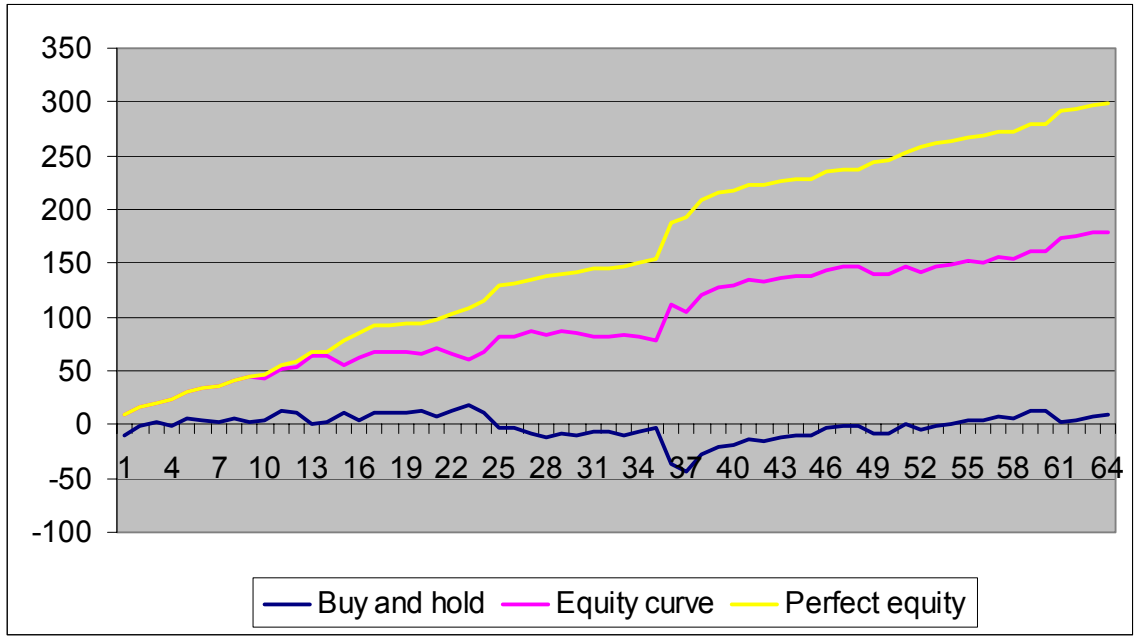

**IGAEM Curve vs. Network 1 Prediction**

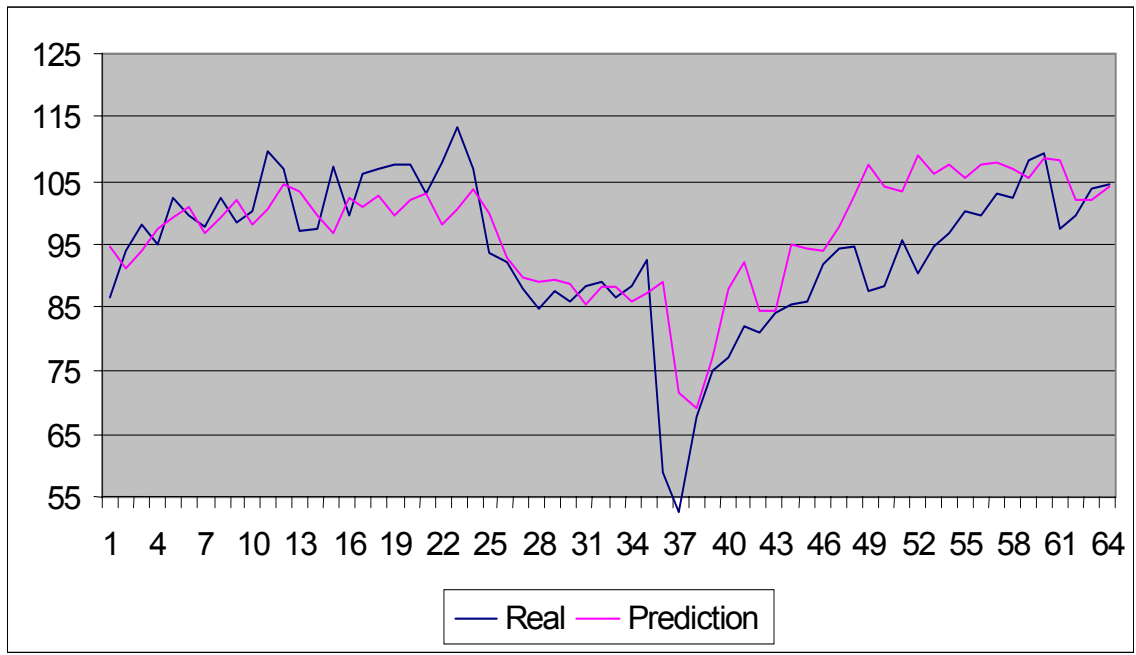

**Network 2 Equity Curve**

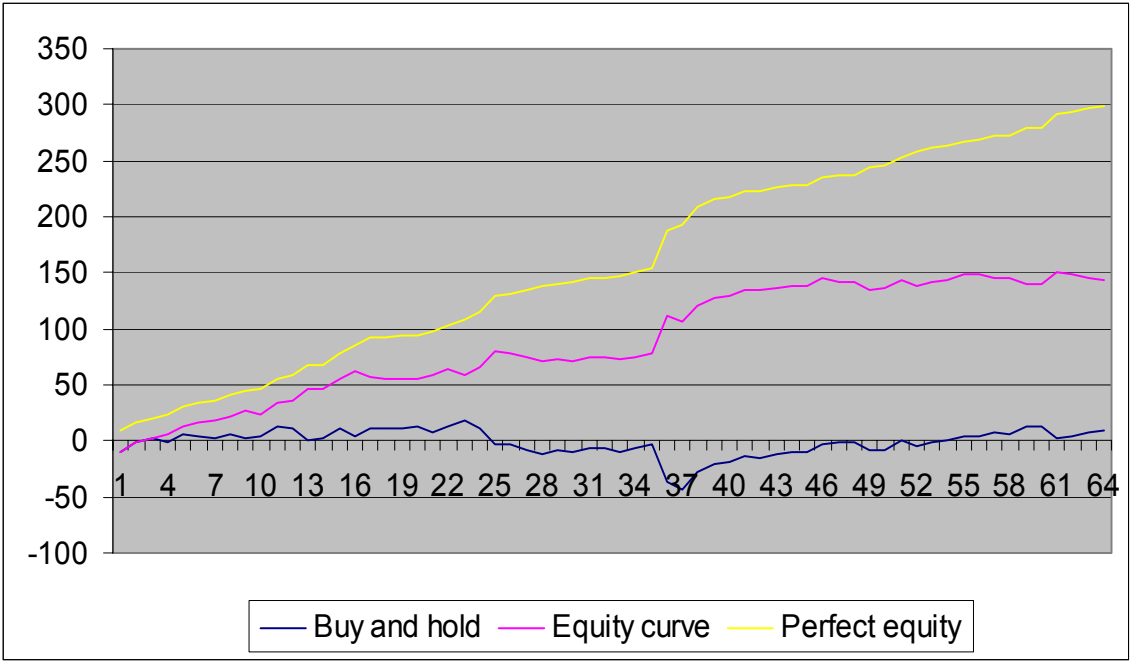

**IGAEM Curve vs. Network 2 Prediction**

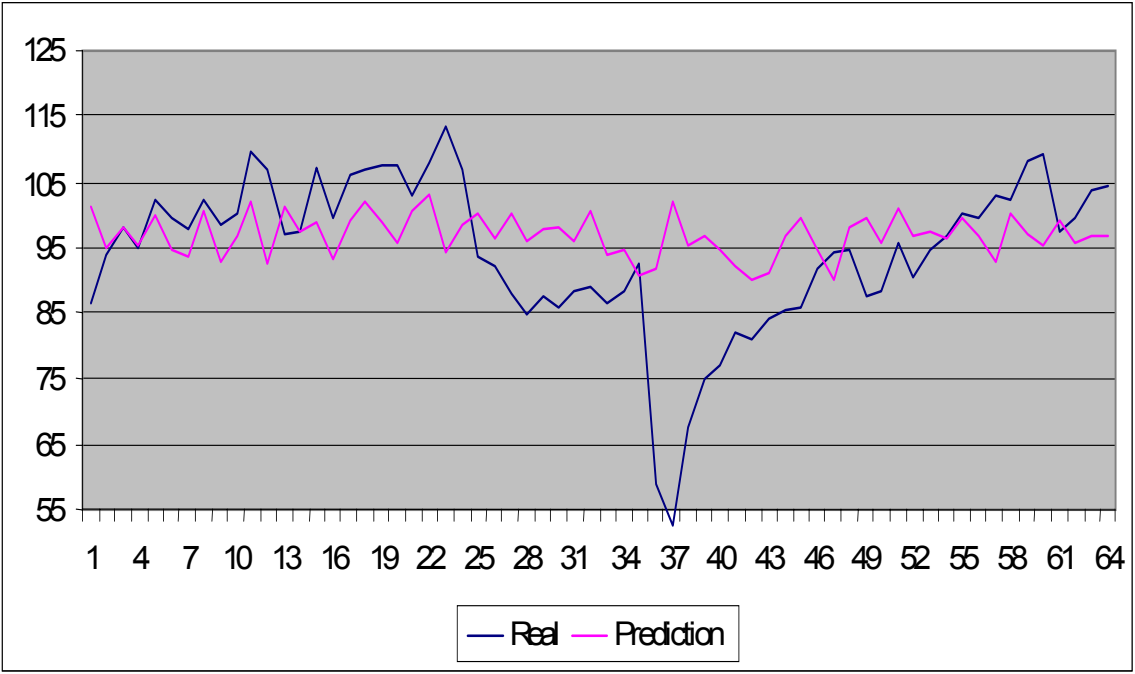

**Network 3 Equity Curve**

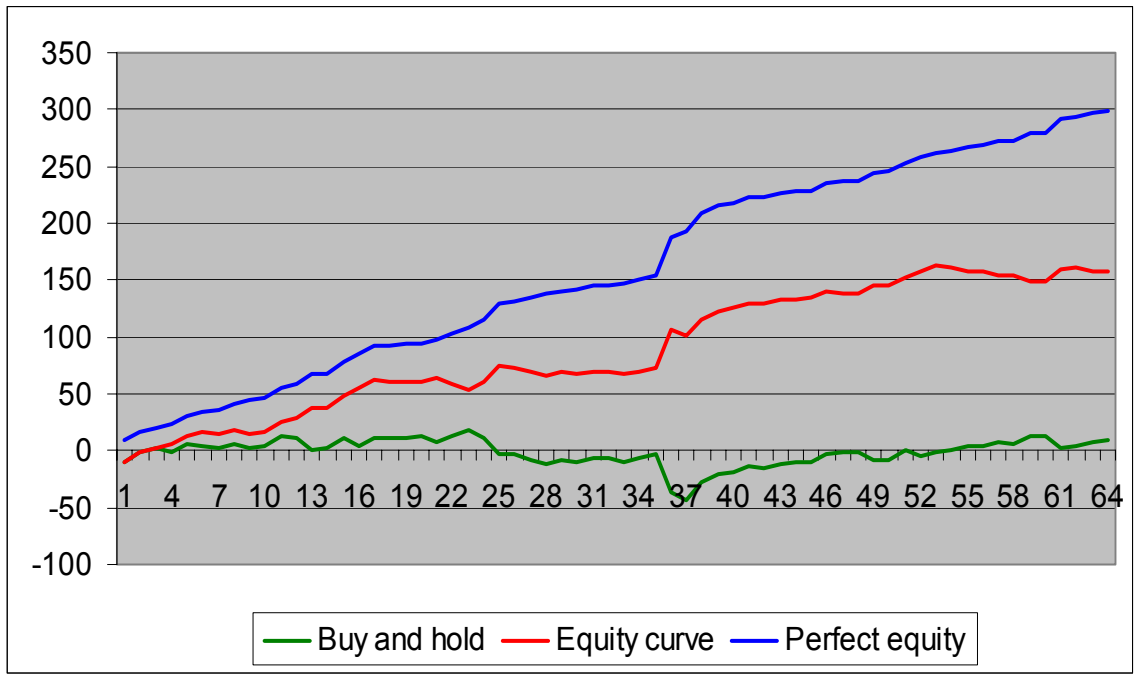

**IGAEM Curve vs. Network 3 Prediction**

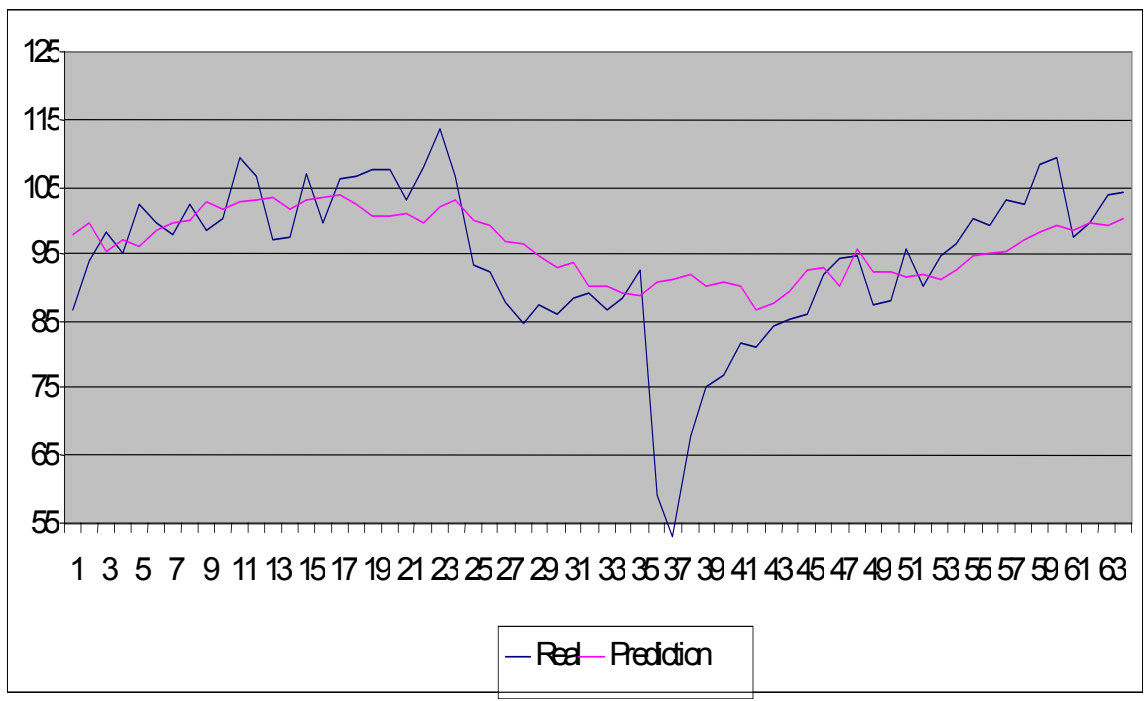

**Network 4 Equity Curve**

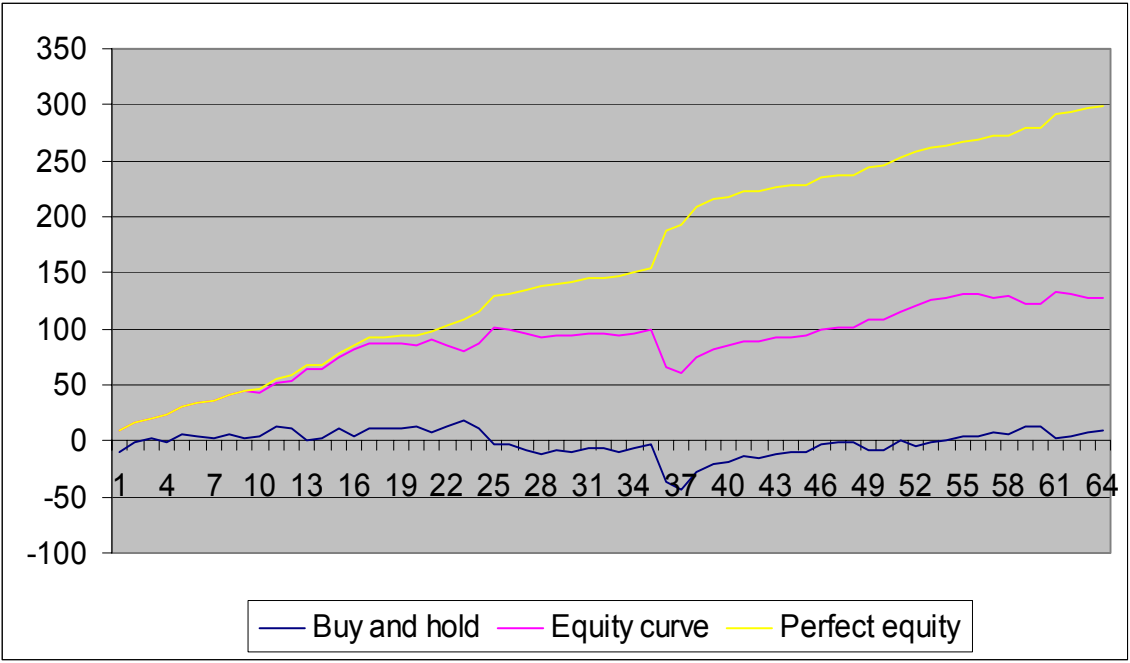

**IGAEM Curve vs. Network 4 Prediction**

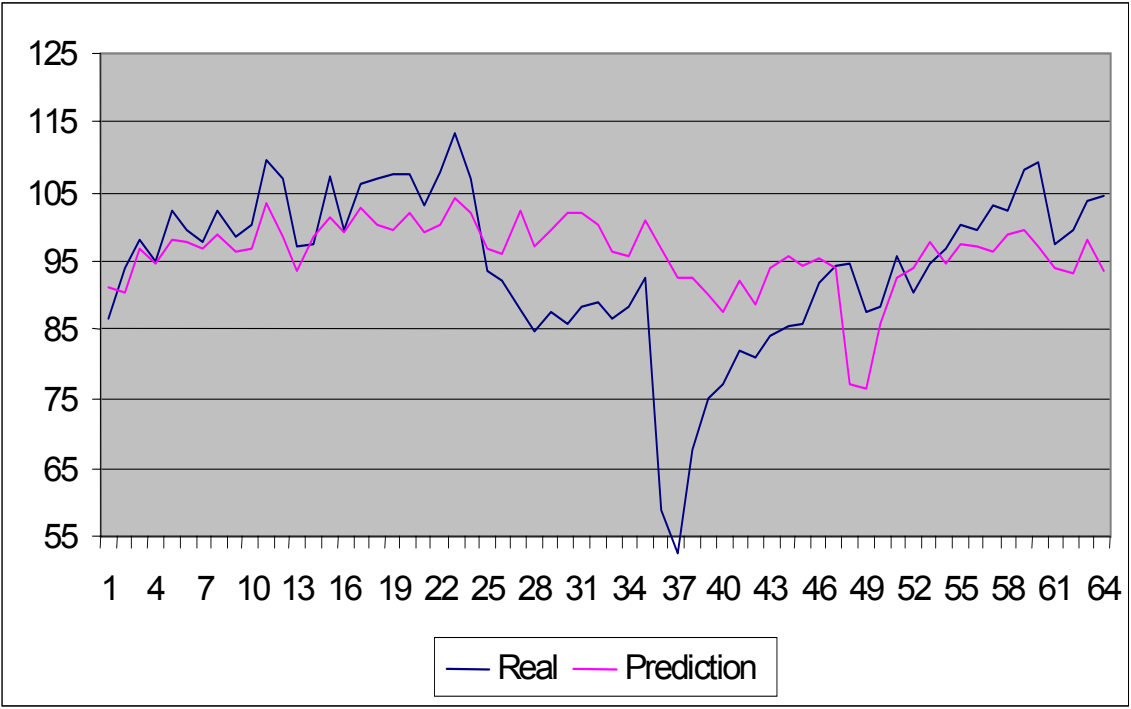

**Network 5 Equity Curve**

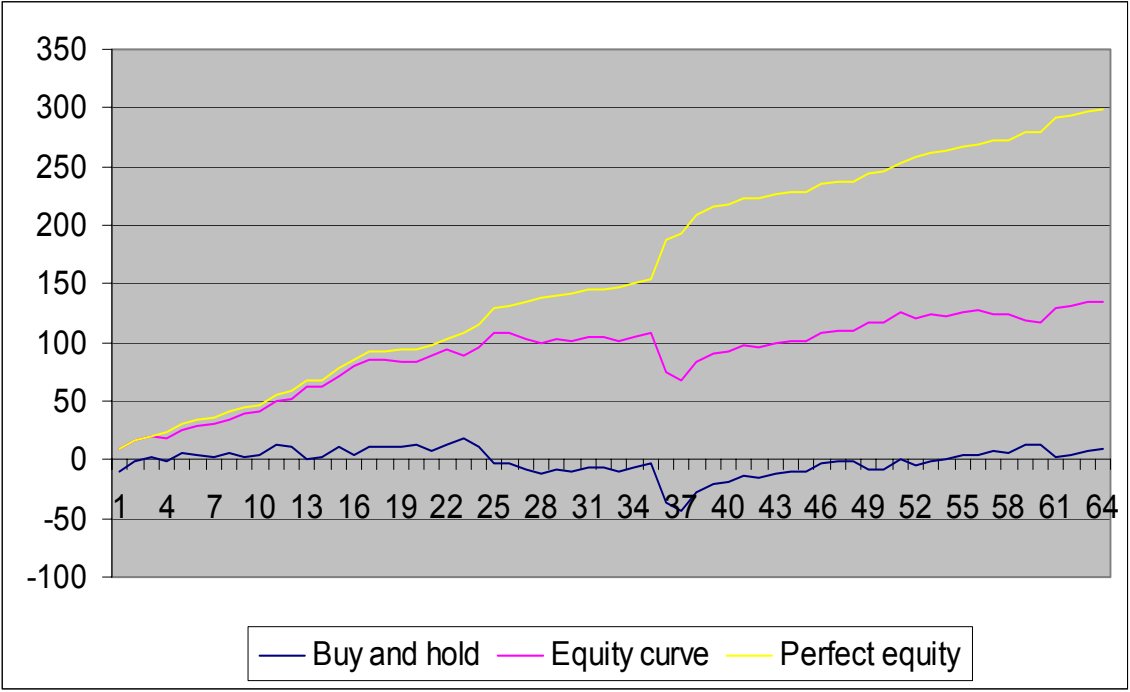

**IGAEM Curve vs. Network 5 Prediction**

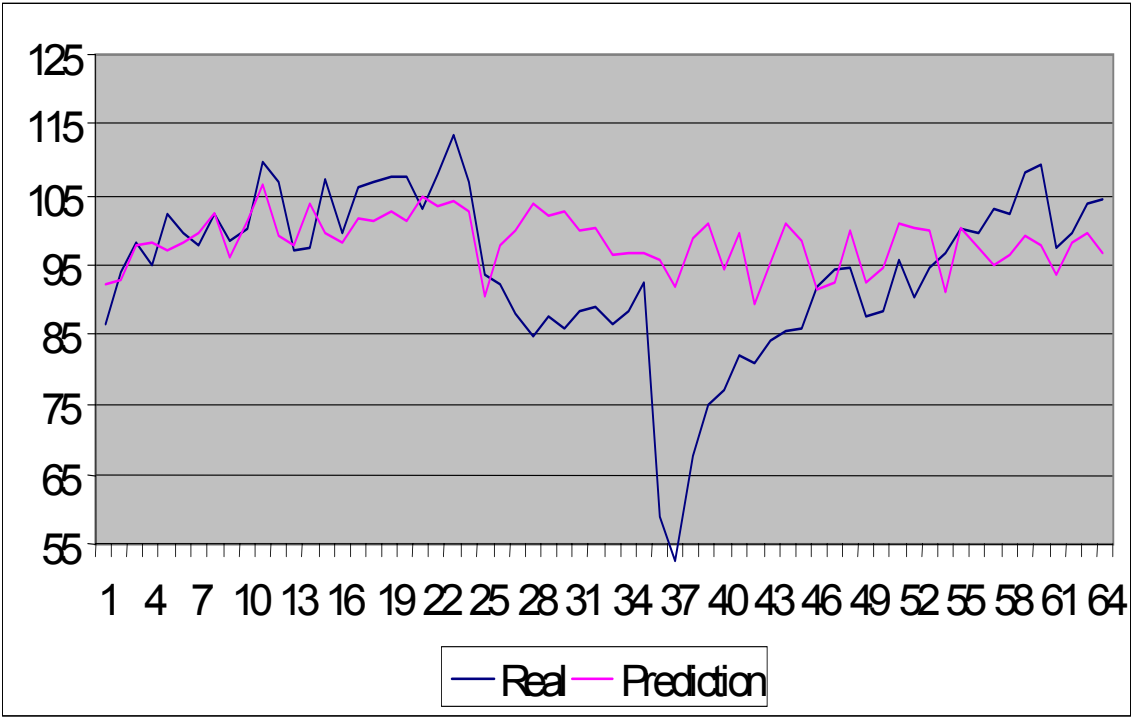

**Network 6 Equity Curve**

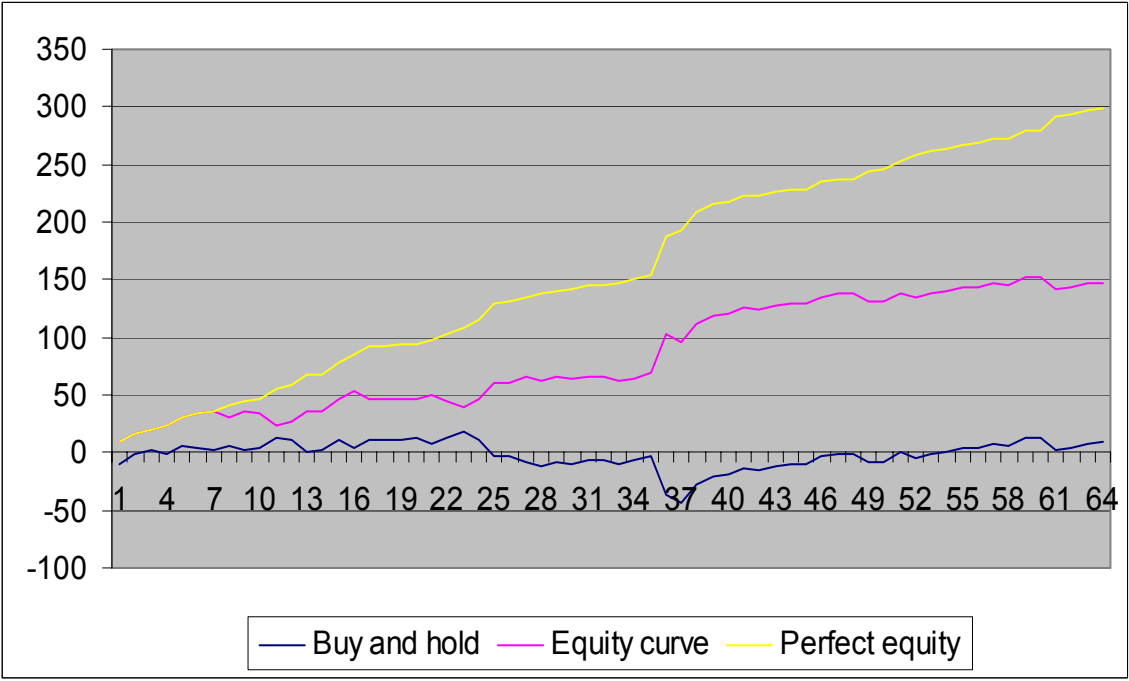

**IGAEM Curve vs. Network 6 Prediction**

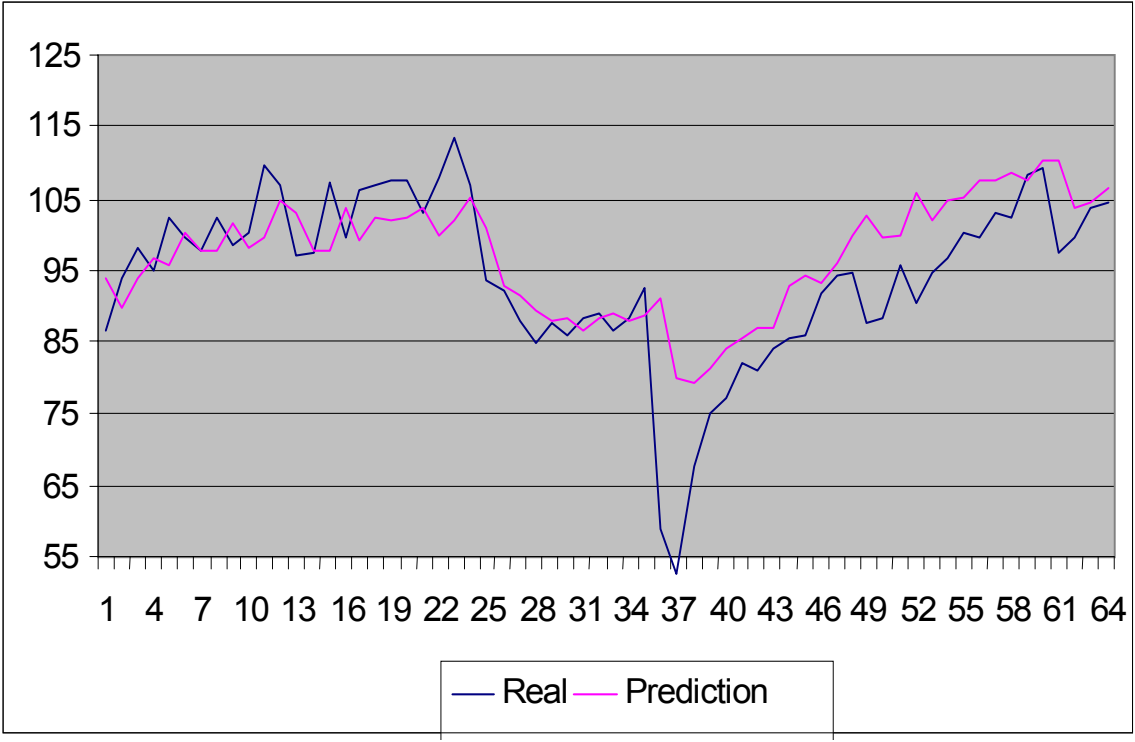

**Network 7 Equity Curve**

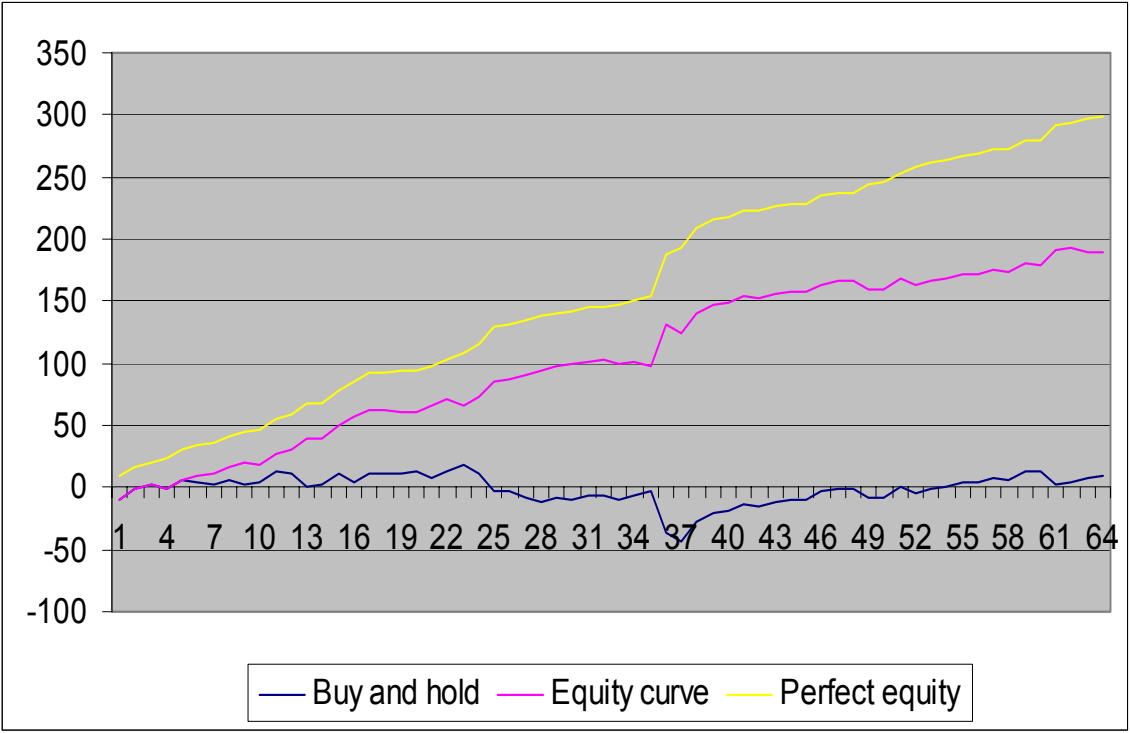

**IGAEM Curve vs. Network 7 Prediction**

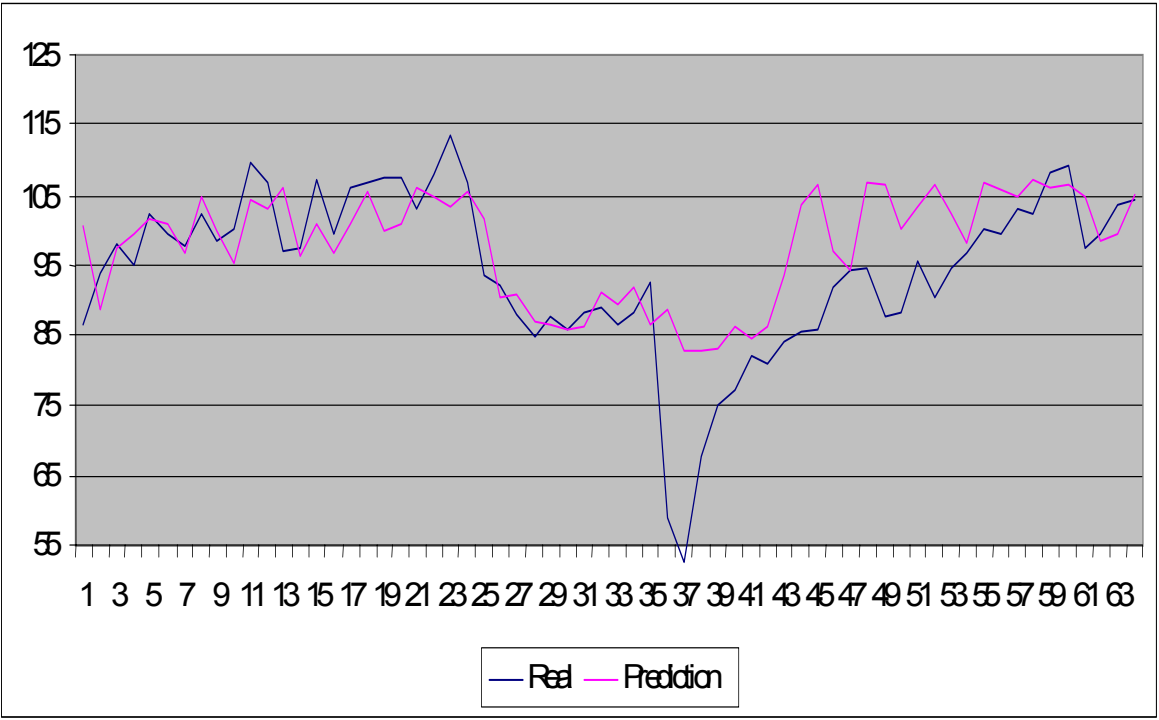

**Network 8 Equity Curve**

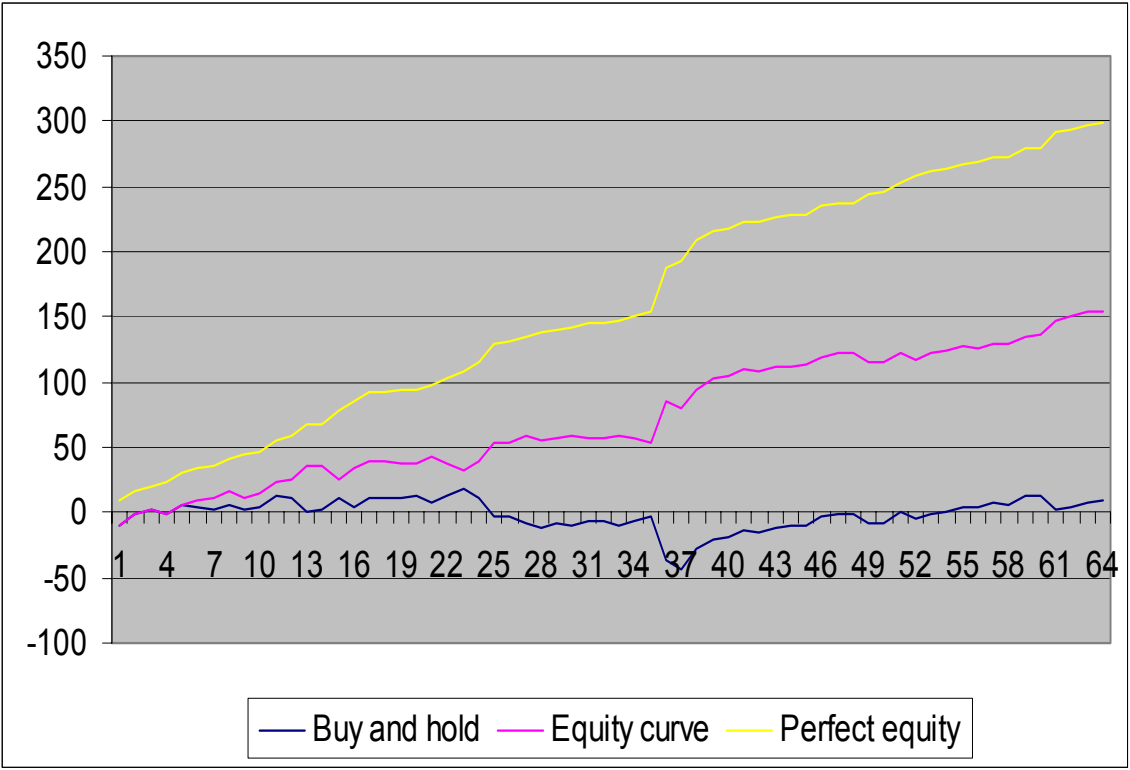

**IGAEM Curve vs. Network 8 Prediction**

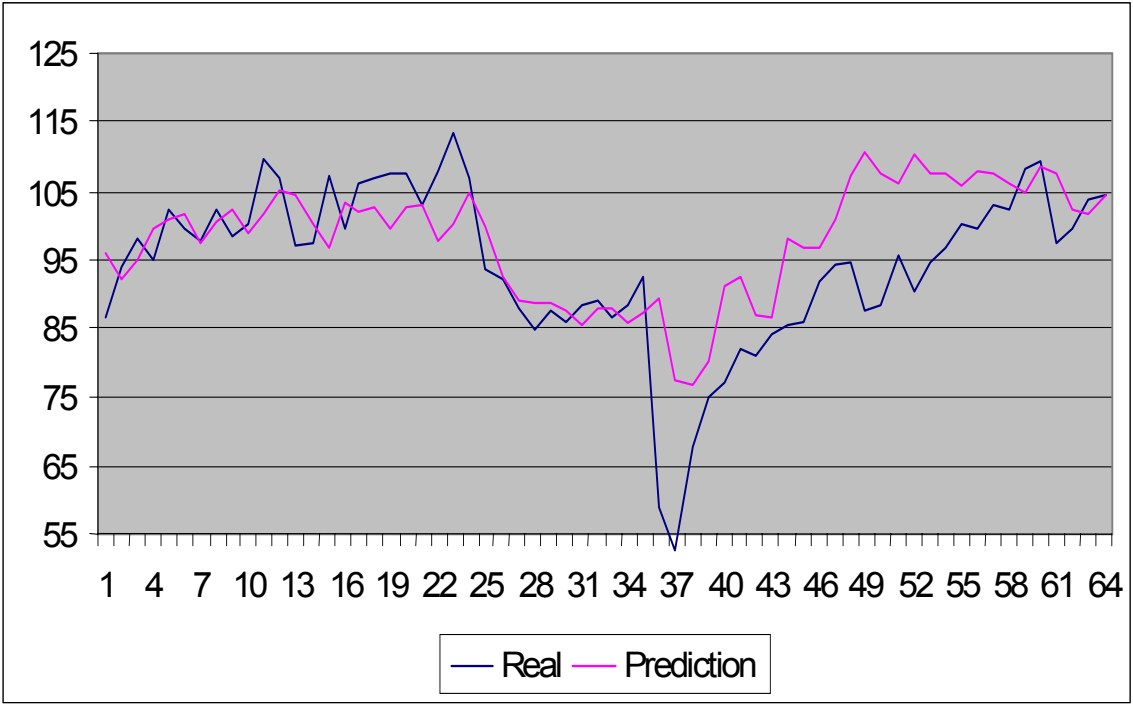